\documentclass{webofc}
\usepackage[varg]{txfonts}   

\usepackage{subcaption}
\begin{document}
\title{Constraints on anomalous couplings of the Higgs boson from pair production searches}
%
%


\author{\firstname{Petr} \lastname{Mandrik}\inst{1,2}\thanks{\email{petr.mandrik@ihep.ru}}
}

\institute{NRC ``Kurchatov Institute'' -- IHEP, Protvino \and Moscow Institute of Physics and Technology, Dolgoprudny}

\abstract{%
The study of the anomalous non-resonant $HH$ production is presented from the point of view of effective field theory (EFT) parameterization.
The searches for
$pp \rightarrow HH \rightarrow \gamma \gamma WW$, 
$pp \rightarrow HH \rightarrow \gamma \gamma bb$ and $pp \rightarrow HH \rightarrow bbbb$ processes by ATLAS experiment at the Large Hadron Collider (LHC) are reproduced.
The selection efficiencies and exclusion limits on the $HH$ production cross sections for different Beyond Standard Model (BSM) EFT benchmark models are obtained using the fast Monte-Carlo simulation of the ATLAS detector.
}
\maketitle
\section{Introduction} \label{intro}
The discovery of the Higgs boson by the Large Hadron Collider (LHC)~\cite{Aad:2012tfa, Chatrchyan:2012xdj} experiments has opened up a
new area of direct searches for physics Beyond Standard Model (BSM) using the Higgs boson as a probe \cite{deFlorian:2016spz, Hou:2017vvp, Zhang:2018nmy, Ilyushin:2019mkp, Capozi:2019xsi, Cao:2015oaa}.
An important test of the Standard Model (SM) is the measurement of Higgs boson pair production. 
In particular, many BSM models predict the existence of heavy particles that can couple to a pair of Higgs bosons \cite{deFlorian:2016spz, Nakamura:2017irk, Englert:2019eyl, Robens:2019kga, Tang:2012pv}.
These particles could appear as a resonant contribution to the invariant mass of the $HH$ system or they may contribute
to Higgs boson pair production through virtual processes and lead to the cross sections for Higgs boson pair production that are significantly greater than the SM prediction.
In this article we use recent results from LHC searches in different final states \cite{Aaboud:2018ftw, Aaboud:2018ewm, Aaboud:2018knk} in order to put limits on anomalous interactions of the Higgs bosons. The study is based on Monte-Carlo (MC) simulation of related processes which allow to take into account differences in SM and BSM Higgs boson pair productions and incorporate the detector effects and reconstruction efficiencies.

The rest part of this paper is organized as follows.
In Section \ref{mc} we describe the theoretical model and MC simulation.
In Section \ref{selection} the event selection and systematic uncertainties are presented.
Finally, the main results are summarized and discussed in Section \ref{results}.

\section{Event generation} \label{mc}
While the BSM physics in Higgs interactions may arise from different sources 
the effective field theory approach (EFT)~\cite{Weinberg:1978kz, Buchmuller:1985jz, Arzt:1994gp}
is used to parameterize observable effects.
In this article we use the following effective Lagrangian (with up to dimension-six operators)~\cite{deFlorian:2016spz, Carvalho:2015ttv} to describe the Higgs boson pair production:
\begin{equation} \label{eq_lagrangian}
  \mathcal{L}_{BSM} = \kappa_{\lambda} \lambda^{SM}_{HHH} v H^3 - \frac{m_t}{v}(\kappa_t H + \frac{c_2}{v} H^2)(\bar{t_L}t_R + h.c.) + \frac{1}{4} \frac{\alpha_S}{3 \pi v}(c_g H - \frac{c_{2g}}{2 v}H^2)G^{\mu \nu}G_{\mu \nu}
\end{equation}
where $v$ - vacuum-expectation value of the Higgs field; 
$\kappa_{\lambda} = \lambda_{HHH} / \lambda_{HHH}^{SM}$ - measure of deviation of Higgs boson trilinear coupling from its SM expectation $\lambda_{HHH}^{SM} = m^2_{H} / 2v^2$; 
$\kappa_{t} = y_{t} / y_{t}^{SM}$ - measure of deviation of Higgs boson trilinear coupling from its SM expectation $y_{t}^{SM} = \sqrt{2} m^2_{t} / v$;
$c_{2}$ - coupling between two Higgs bosons and two top quarks;
$c_{g}$ - coupling between one Higgs bosons and two gluons;
$c_{2g}$ - coupling between two Higgs bosons and two gluons.

\begin{figure}
  \centering
  \begin{subfigure}[t]{0.40\textwidth}
    \centering
    \raisebox{-0.5\height}{ \includegraphics[width=\linewidth,clip]{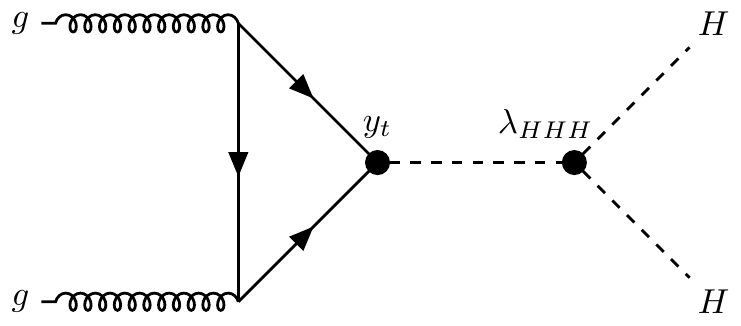} }
  \end{subfigure}
  \begin{subfigure}[t]{0.40\textwidth}
    \centering
    \raisebox{-0.425\height}{ \includegraphics[width=\linewidth,clip]{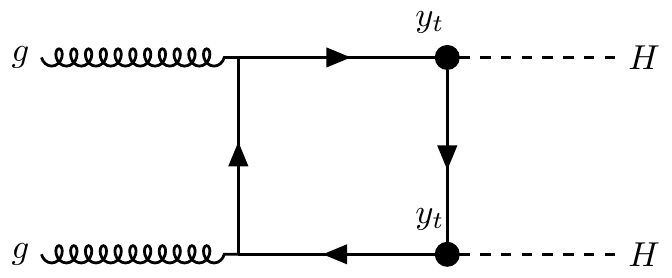} }
  \end{subfigure}
  \\
  \begin{subfigure}[t]{0.30\textwidth}
    \centering
    \raisebox{-0.5\height}{ \includegraphics[width=\linewidth,clip]{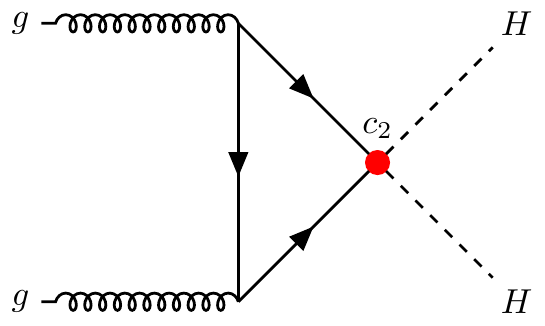} }
  \end{subfigure}
  \begin{subfigure}[t]{0.20\textwidth}
    \centering
    \raisebox{-0.5\height}{ \includegraphics[width=\linewidth,clip]{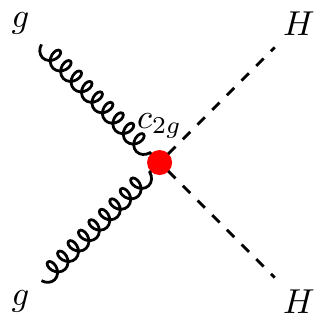} }
  \end{subfigure}
  \begin{subfigure}[t]{0.30\textwidth}
    \centering
    \raisebox{-0.5\height}{ \includegraphics[width=\linewidth,clip]{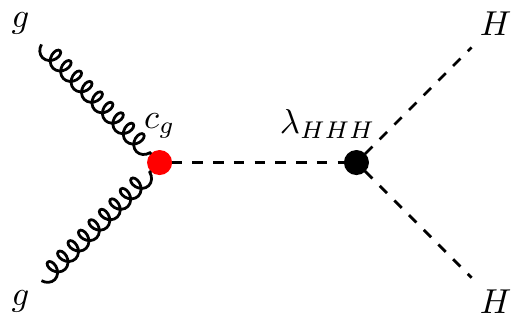} }
  \end{subfigure}

  \caption{ Feynman diagrams for leading-order Higgs boson pair production. 
            Top diagrams correspond to SM-like processes. The bottom diagrams correspond to pure BSM processes. }
  \label{fcnc_decays}       
\end{figure}

As base events, we use the 12 benchmarks defined in \cite{Carvalho:2015ttv, Carvalho:2016rys} and for each of them we simulate $500 \cdot 10^3$ events. 
The usage of a manageably small set of benchmark points allow to represent the volume of the unexplored parameter space. 
The list of benchmark hypotheses is provided in Table \ref{table_fcnc}.

\begin{table*}[h]
  \centering
\caption{Parameter values of non-resonant BSM benchmark hypotheses of Higgs boson pair production. The first column correspond to the SM sample, while the next 12 correspond to the benchmark hypotheses ~\cite{Carvalho:2015ttv, Carvalho:2016rys}. Note that (1, 1, 0, 0, 0) values of $(\kappa_\lambda$, $\kappa_t$, $c_{2}$, $c_{g}$, $c_{2g})$ set corresponds to the SM production. \label{table_fcnc}}
  \begin{tabular}{ l | c c c c c c c c c c c c }
  \hline
                     & 1 &   2 &    3 &    4 &   5 &   6 &   7 &    8 &   9 &   10 &  11 &   12 \\
  \hline
  $\kappa_\lambda$   &  7.5 &  1.0 &  1.0 & -3.5 &  1.0 &  2.4 &  5.0 & 15.0 &  1.0 & 10.0 &  2.4 & 15.0 \\
  $\kappa_t$         &  1.0 &  1.0 &  1.0 &  1.5 &  1.0 &  1.0 &  1.0 &  1.0 &  1.0 &  1.5 &  1.0 &  1.0 \\
  $c_{2}$            & -1.0 &  0.5 & -1.5 & -3.0 &  0.0 &  0.0 &  0.0 &  0.0 &  1.0 & -1.0 &  0.0 &  1.0 \\
  $c_{g}$            &  0.0 & -0.8 &  0.0 &  0.0 &  0.8 &  0.2 &  0.2 & -1.0 & -0.6 &  0.0 &  1.0 &  0.0 \\
  $c_{2g}$           &  0.0 &  0.6 & -0.8 &  0.0 & -1.0 & -0.2 & -0.2 &  1.0 &  0.6 &  0.0 & -1.0 &  0.0 \\
  \hline
  \end{tabular}
\end{table*}

The FeynRules~\cite{Alloul:2013bka} implementation of the Lagrangian (\ref{eq_lagrangian})
is interfaced with {\scshape MG5\_}a{\scshape MC@NLO}~2.4.2~\cite{Alwall:2014hca} package using the UFO module~\cite{Degrande:2011ua}.
All generated events are processed with {\scshape Pythia}~8.230~\cite{Sjostrand:2014zea} for showering, hadronization and the underlying event description.
The {\scshape NNPDF3.0} \cite{Ball:2014uwa} PDF sets are used.
The detector simulation is performed with the fast simulation
tool~{\scshape Delphes}~3.4.2~\cite{deFavereau2014}
using the corresponding detectors parameterization cards. No additional
pileup interactions are added to the simulation.

\section{Event selection and systematic uncertainties} \label{selection}
The event selections from $HH \rightarrow \gamma \gamma bb$~\cite{Aaboud:2018ftw}, $HH \rightarrow \gamma \gamma WW$~\cite{Aaboud:2018ewm} 
and $HH \rightarrow bbbb$~\cite{Aaboud:2018knk} searches are reproduced in order to accurately estimate the efficiency for the BSM benchmark hypotheses.
In all analyses we use anti-$k_T$ jets reconstructed with a radius parameter $R = 0.4$.

\subsection{ $HH \rightarrow \gamma \gamma WW$ channel }
For the search of $\gamma \gamma \ell \nu j j$ final state following pre-selected objects are used: 
photons with $p_T > 25$ Gev, $|\eta| < 2.37$ (excluding $1.37 < |\eta| < 1.52$), $I_{rel} < 0.05$,
electrons with $p_T > 10$ Gev, $|\eta| < 2.37$ (excluding $1.37 < |\eta| < 1.52$), $I_{rel} < 0.05$,
muons with $p_T > 10$ Gev, $|\eta| < 2.7$, $I_{rel} < 0.05$ and
jets with $p_T > 25$ Gev, $|\eta| < 2.5$. 
Events are selected using a diphoton trigger, which requires two photon candidates, one with transverse energy $E_T > 35$ GeV and second with $E_T > 25$ GeV.

An overlap removal procedure is performed in the following order: electrons with $\Delta R(e, \gamma) < 0.4$ are removed, jets with
$\Delta R(jet, \gamma) < 0.4$ or $\Delta R(jet, e) < 0.2$ are removed, electrons with $\Delta R(e, jet) < 0.4$ are removed, muons
with $\Delta R(\mu, \gamma) < 0.4$ or $\Delta R(\mu, jet) < 0.4$ are removed.

The events are required to contain at least two jets, no b-tagged jets, at least one charged lepton ($e$ or $\mu$).
The transverse momentum of the diphoton system of leading and sub-leading $E_T$ photons is required to be larger than 100 GeV and
the mass of diphoton system is required to be $121.7 \text{ GeV} < m_{\gamma \gamma} < 128.5$ GeV.
The leading (sub-leading) photon candidate is required to satisfy $E_T /m_{\gamma \gamma} > 0.35$ (0.25).

\subsection{ $HH \rightarrow \gamma \gamma bb$ channel }
Object pre-selections and trigger are the same as in $HH \rightarrow \gamma \gamma WW$ searches.
In additional any jets that are within $\Delta R = 0.4$ of an
isolated photon candidate or within $\Delta R = 0.2$ of an isolated electron candidate are discarded.

The events are required to contain exactly two b-tagged jets with $p_T > 100$ GeV of the leading and $p_T > 30$ GeV of the second jet and
at least two photons. The leading (sub-leading) photon candidate is required to satisfy $E_T /m_{\gamma \gamma} > 0.35$ (0.25).
The dijet invariant mass is required to be within mass window of $90$ GeV $< m_{jj} < 140$ GeV and the diphoton invariant mass is required to fall within $105$ GeV $< m_{jj} < 160$ GeV.

\subsection{ $HH \rightarrow bbbb$ channel } 
The trigger effects are taken into account requiring the events to feature either one $b-$tagged jet with $p_T > 225$ GeV,
or two $b-$tagged jets with $p_T > 55$ GeV.
The events are required to contain at least four b-tagged jets with $p_T > 40$ GeV and $|\eta| < 2.5$.
The b-tagging jets are paired to construct two Higgs boson candidates.
The leading (sub-leading) Higgs boson candidates should have 
$\Delta R_{jj, lead} \cdot m_{4j} \in \big[ 360 \text{ GeV} - 0.5 \cdot m_{4j},  653 \text{ GeV} + 0.475 \cdot m_{4j} \big]$
($\Delta R_{jj, subl} \cdot m_{4j} \in \big[ 235 \text{ GeV},  875 \text{ GeV} + 0.35 \cdot m_{4j} \big]$)
if $m_{4j} < 1250$ and $\Delta R_{jj, lead} \in [0, 1]$ ($\Delta R_{jj, subl} \in [0, 1]$) otherwise.
The leading Higgs boson candidate is defined as the candidate with the highest scalar sum of jet $p_T$. 
In case of more than two Higgs boson candidates satisfying these requirements the pairing with the smallest value of mass disbalance $D_{HH}$ is chosen:
\begin{equation}
D_{HH} = \frac{|110 \cdot m_{2j}^{lead} - 120 \cdot m_{2j}^{subl}|}{\sqrt{110^2 + 120^2}}
\end{equation}
Selected event should contain the leading Higgs boson candidate
with $p_T^{lead} > 0.5 m_{4j} - 103$ GeV
and sub-leading Higgs boson candidates 
with $p_T^{subl} > 0.33 m_{4j} - 73$ GeV.
The requirement $|\Delta \eta| < 1.5$ is placed on the pseudorapidity difference between the two Higgs boson candidates.
A further requirement on the Higgs boson candidates masses is applied:
\begin{equation}
  \Big( \frac{ m_{2j}^{lead} - 120 \text{ GeV} }{ 0.1 m_{2j}^{lead} } \Big)^2
+ \Big( \frac{ m_{2j}^{subl} - 110 \text{ GeV} }{ 0.1 m_{2j}^{subl} } \Big)^2 < 2.56
\end{equation}
Finally, all possible hadronically decaying top-quark candidates are built from combinations of three jets
of which one must be a constituent of a Higgs boson candidate and the other have $p_T > 10$ Gev and $|\eta| < 2.5$.
Event is vetoed in the final selection if it contain a top-quark candidate with:
\begin{equation}
  \Big( \frac{ m_{W} - 80 \text{ GeV} }{ 0.1 m_{W} } \Big)^2
+ \Big( \frac{ m_{t} - 173 \text{ GeV} }{ 0.1 m_{t} } \Big)^2 < 2.25
\end{equation}

\subsection{ Selection efficiencies } 
\begin{table*}[t]
  \centering
\caption{ Combined acceptance and efficiency in \% of different benchmark points. \label{table_eff}}
  \begin{tabular}{ c | c c c }
  \hline
  Benchmark         & $HH \rightarrow \gamma \gamma WW$ & $HH \rightarrow \gamma \gamma bb$ & $HH \rightarrow bbbb$ \\
  \hline
  SM (Geant4) \cite{Aaboud:2018ftw, Aaboud:2018ewm, Aaboud:2018knk}      &  8.5 &  5.8  &  1.6  \\ 
 SM (Delphes) & 6.5 $\pm$ 1.7 & 5.5 $\pm$ 1.5 & 1.3 $\pm$ 0.4 \\ 
 \hline 
1 & 6.3 $\pm$ 1.8 & 4.3 $\pm$ 1.3 & 1.1 $\pm$ 0.3 \\ 
2 & 4.9 $\pm$ 1.5 & 8.4 $\pm$ 2.7 & 1.5 $\pm$ 0.5 \\ 
3 & 5.6 $\pm$ 1.6 & 6.5 $\pm$ 1.9 & 1.3 $\pm$ 0.4 \\ 
4 & 5.7 $\pm$ 1.6 & 4.9 $\pm$ 1.4 & 1.1 $\pm$ 0.3 \\ 
5 & 5.3 $\pm$ 1.5 & 5.6 $\pm$ 1.7 & 1.3 $\pm$ 0.4 \\ 
6 & 5.5 $\pm$ 1.5 & 5.5 $\pm$ 1.6 & 1.1 $\pm$ 0.3 \\ 
7 & 5.5 $\pm$ 1.5 & 4.4 $\pm$ 1.3 & 1.1 $\pm$ 0.3 \\ 
8 & 5.2 $\pm$ 1.5 & 5.3 $\pm$ 1.6 & 1.1 $\pm$ 0.3 \\ 
9 & 7.0 $\pm$ 2.1 & 10.0 $\pm$ 3.1 & 1.9 $\pm$ 0.6 \\ 
10 & 5.7 $\pm$ 1.6 & 4.5 $\pm$ 1.3 & 1.1 $\pm$ 0.3 \\ 
11 & 6.5 $\pm$ 1.8 & 4.3 $\pm$ 1.3 & 1.2 $\pm$ 0.3 \\ 
12 & 5.4 $\pm$ 1.5 & 4.9 $\pm$ 1.4 & 1.1 $\pm$ 0.3 \\ 
\hline
  \end{tabular}
\end{table*}

The systematic uncertainty in the photon identification and isolation is 3\% in total signal yelds \cite{Aaboud:2018yqu}, in the integrated luminosity is 2.1\%, in trigger efficiency is 0.4\% for $HH \rightarrow \gamma \gamma WW$ and $HH \rightarrow \gamma \gamma bb$ analyses and 2.5\% in $HH \rightarrow bbbb$ searches.
In all analyses the theoretical uncertainty from the renormalization and factorization scale is determined by varying
these scales between 0.5 and 2 times their nominal value while keeping their ratio between 0.5 and 2 \cite{deFlorian:2016spz}.
PDF uncertainty is determined by taking the root mean square of the variation when using different replicas of the default PDF set~\cite{Butterworth:2015oua}.
The impact in total signal yelds of the jet energy scale systematic uncertainty is estimated following the prescription in \cite{Aad:2014bia}.

Finally, the selection efficiencies for different benchmark points obtained from the MC simulation are
given in Table~\ref{table_eff}. The uncertainties are combined using summation in quadrature.
In the following statistical analysis the correlations between uncertainties for signal and background are neglected.

\section{Results and conclusions} \label{results}
Bayesian inference is used to derive the posterior probability based on the number of selected events
where the expected number of signal events is from our modeling and the observed number of events and expected number of background events with uncertainties are taken from the corresponding experimental results \cite{Aaboud:2018ftw, Aaboud:2018ewm, Aaboud:2018knk}. 
The exclusion limits at 95\% C.L. on $HH$ production crossection times branchings for different benchmark models are given in Table \ref{table_res}.
The limits on $\sigma_{pp \rightarrow HH \rightarrow \gamma \gamma WW}$ obtained for the first time. 
The limits on $\sigma_{pp \rightarrow HH \rightarrow \gamma \gamma bb}$
and $\sigma_{pp \rightarrow HH \rightarrow bbbb}$ are
comparable with results from \cite{Sirunyan:2018tki, Sirunyan:2018qca} and \cite{Sirunyan:2018iwt} respectly.
In order to put limits on $HH$ production crossection based on the combination of analyses we use 
$B(H \rightarrow bb) = 0.5824$, $B(H \rightarrow \gamma \gamma) = 2.27 \times 10^{-3}$, $B(H \rightarrow WW) = 0.2137$ 
with additional theoretical uncertainties $0.65\%$, $1.73\%$ and $1\%$ respectively \cite{deFlorian:2016spz}.
The obtained limits on $\sigma_{pp \rightarrow HH}$ are also can complement the results of the combination analysis shown in supplemental material of \cite{CMS-PAS-HIG-17-030}.
Future improvements can be obtained from combinations with other ATLAS $HH$ searches.
As the next step, the resulting exclusion bounds could be used in order to constrain BSM models, mapped to EFT, as described in \cite{Carvalho:2017vnu}.

\begin{table*}[h]
  \centering
\caption{  Observed 95\% CL upper limits in fb on the $HH$ production crossection times branchings obtained for different benchmark models and the limits on the crossections from combination. } \label{table_res}
  \begin{tabular}{ c | c c c | c }
  \hline
  Benchmark         & $\sigma_{pp \rightarrow HH \rightarrow \gamma \gamma WW}$ & $\sigma_{pp \rightarrow HH \rightarrow \gamma \gamma bb}$ & $\sigma_{pp \rightarrow HH \rightarrow bbbb}$ & $\sigma_{pp \rightarrow HH}$  \\ \hline 
1 & 14.6 & 2.4 & 1437.0 & 1372.6  \\ 
2 & 18.6 & 1.2 & 1101.9 & 1149.1  \\ 
3 & 16.5 & 1.6 & 1252.5 & 1094.7  \\ 
4 & 16.1 & 2.2 & 1491.1 & 1316.7  \\ 
5 & 17.3 & 1.9 & 1295.3 & 1248.7  \\ 
6 & 16.9 & 1.9 & 1462.5 & 1223.9  \\ 
7 & 16.6 & 2.4 & 1442.8 & 1365.1  \\ 
8 & 17.7 & 2.0 & 1484.8 & 1247.8  \\ 
9 & 13.1 & 1.0 & 868.9 & 866.1    \\ 
10 & 16.0 & 2.4 & 1481.2 & 1373.7 \\ 
11 & 14.1 & 2.4 & 1390.7 & 1366.5 \\ 
12 & 17.1 & 2.2 & 1510.9 & 1327.0 \\ 
\hline
  \end{tabular}
\end{table*}

\section{Acknowledgments}
I would like to thank S.~Slabospitskii and V.~Kachanov for useful discussions.
%
\bibliography{pmandrik_bib_file.bib}
%
%

\end{document}